\documentclass[reprint,nofootinbib,aps,superscriptaddress,amsmath,amssymb,onecolumn,11pt]{revtex4-2}

\usepackage{graphicx}
\usepackage{dcolumn}% Align table columns on decimal point
\usepackage{bm}% bold math
\usepackage{braket}
\usepackage{breqn}
\usepackage{mathrsfs}
\usepackage{color}

\usepackage[colorlinks=true,linkcolor=red,citecolor=blue]{hyperref}

\begin{document}

\title{Nonlinear instability and scalar clouds of spherical exotic compact objects in scalar-Gauss-Bonnet theory}

\author{Shao-Jun Zhang}
\email{sjzhang@zjut.edu.cn}
\affiliation{Institute for Theoretical Physics and Cosmology$,$ Zhejiang University of Technology$,$ Hangzhou 310032$,$ China}
\affiliation{United Center for Gravitational Wave Physics$,$ Zhejiang University of Technology$,$ Hangzhou 310032$,$ China}
\date{\today}

\begin{abstract}
	\indent In this work, we present a new type of scalar clouds supported by spherically symmetric horizonless compact objects in the scalar-Gauss-Bonnet theory. Unlike the previous spontaneous scalarization that is triggered by the tachyonic instability, our scalarization arises from a nonlinear instability that is non-spontaneous. We explore two types of boundary conditions for the scalar field at the surface of the compact objects and find an infinite countable set of scalar clouds characterized by the number of nodes for both cases. Our study demonstrates that boundary conditions have a significant impact on the formation of scalar clouds. Specifically, for the Dirichlet boundary condition, scalarization is more likely to occur for compact objects with medium radii and becomes harder for ultra-compact and large ones. Conversely, for the Robin boundary condition, scalarization is easier for more compact objects.
\end{abstract}

%\pacs{}

\maketitle
\section{Introduction}

Black holes (BHs) are one of the most intriguing predictions of general relativity (GR) and have attracted extensive attention in recent decades due to their significance. The detection of gravitational waves \cite{LIGOScientific:2016aoc} and the imaging of BHs \cite{EventHorizonTelescope:2019dse} provide compelling evidence of their existence. However, several challenging issues, such as the information loss paradox \cite{Mathur:2009hf} and the emergence of unphysical curvature singularities \cite{penrose1965Gravitational}, remain unresolved. To address these issues, physicists have proposed several alternatives to BHs, collectively known as exotic compact objects (ECOs) \cite{kaup1968KleinGordon,Ruffini:1969qy,Liebling:2012fv,Mazur:2001fv,Mazur:2004fk,Lunin:2002qf,Mathur:2005zp,Brustein:2016msz,Maggio:2017ivp}, whose mass and compactness can closely match those of BHs, making them potential candidates for explaining astronomical phenomena attributed to BHs. Theoretically, BHs and ECOs may be seen as distinct "species" of compact objects, which could co-exist in our universe. Notably, some recent gravitational wave mergers have the possibility to involve ECOs \cite{CalderonBustillo:2020fyi}. Nevertheless, our understanding of ECOs is significantly less than that of BHs. The exploration of ECOs' physical properties may help us to gain a deeper understanding of BHs and quantum features of gravity. With advancements in astronomical observations, particularly in the gravitational waves channel, we expect to access the near-horizon (surface) structure of BHs (ECOs) soon and differentiate between them. For a review, see \cite{Cardoso:2019rvt,Maggio:2021ans}. 

A generic feature of ECOs is the replacement of the classical absorbing horizon by a reflective surface. A simple way to model this feature is proposed in \cite{Maggio:2017ivp}, in which spacetime geometry is only modified at some microscopic scale near the would-be classical horizon while exterior geometry is still described by the usual Kerr metric. Without causing confusion, we will refer to this model as horizonless compact objects in the following. In the framework of GR, one of the most important properties of BHs is given by the no-hair theorem, which states that asymptotically flat static BHs are uniquely characterized by only two physical parameters, their mass and angular momentum, and cannot carry scalar or other types of hair \cite{Bekenstein:1998aw,robinson2004Four,Chrusciel:2012jk}. Therefore, it is interesting to see how modification of the structure close to the would-be horizon affects this theorem and whether ECOs can carry additional hairs. 

Interestingly, it is found that the no-hair theorem applies not only to BHs but also to ECOs in general. In GR, it is found that spherically symmetric horizonless compact objects with a perfectly reflecting surface cannot support minimally coupled scalar, vector or tensor clouds \cite{Hod:2016vkt,Bhattacharjee:2017huz}. The theorem also holds when the scalar field is non-minimally coupled to the Ricci scalar \cite{Hod:2017vij,Hod:2017ehd}. However, if the objects begin to rotate, the situation changes and the no-hair theorem no longer holds. It has been discovered that scalar clouds, massless or massive, can form for rapidly and ultra-spinning horizonless compact objects with a perfectly reflecting surface at certain discrete radii \cite{Hod:2017wks,Hod:2017eld}. Physically, the formation of scalar clouds in this case can be understood from the mechanism of superradiant instability. Namely, with the absorbing horizon replaced by a reflecting surface, spinning ECOs in general suffer from superradiant instability under scalar perturbations \cite{Maggio:2017ivp}, which results in the formation of scalar clouds. A similar phenomenon has already been observed in the BH scenario \cite{Hod:2012px,Herdeiro:2014goa}, but there only massive scalar cloud/hair can be formed.

Another physical mechanism that may endow ECOs with scalar cloud is the tachyonic instability. This phenomenon is inspired by the spontaneous scalarization of BHs in certain modified gravity theories \cite{Doneva:2017bvd,Silva:2017uqg,Antoniou:2017acq,Cunha:2019dwb,Herdeiro:2018wub,Dima:2020yac} (see also \cite{Doneva:2022ewd} for a recent review). The Einstein-scalar-Maxwell (EsM) theory and the scalar-Gauss-Bonnet (sGB) theory are two well-known models of spontaneous scalarization, in which the scalar field is non-minimally coupled to the Maxwell term and the Gauss-Bonnet term, respectively. In both models, it is found that the scalar field perturbation of spherically symmetric horizonless compact objects with perfectly reflecting surface acquires a negative effective mass square, triggering the tachyonic instability, which results in the formation of scalar clouds \cite{Peng:2019qrl,Peng:2019cmm}.

Recently, A new physical mechanism that could induce scalar hair on BHs in EsM or sGB has been discovered \cite{Blazquez-Salcedo:2020nhs,Doneva:2021tvn}, named as nonlinear instability. Different from the tachyonic instability, this type of instability occurs only when the scalar field perturbation becomes relatively large and scalar hair formed are thus non-spontaneous. Motivated by these works, we investigate whether ECOs can also support scalar clouds through this mechanism. This is the main goal of this paper.

The paper is organized as follows. In the next section, we will give a brief review of our model. In Sec. III, we perform time evolution of scalar perturbations of spherically symmetric horizonless compact objects with Dirichlet boundary condition and show the occurrence of the nonlinear instability. In Sec. IV, we construct the scalar clouds with Dirichlet boundary condition induced by the nonlinear instability. In Sec. V, we consider Robin boundary condition to show the influences of different boundary conditions on the formation of scalar clouds. The last section is the Summary and Discussions.

\section{The model}

We consider the scalar-Gauss-Bonnet (sGB) theory with the action \cite{Antoniou:2017acq,Doneva:2017bvd,Silva:2017uqg,Cunha:2019dwb}
\begin{eqnarray}
	S= \int d^4 x \sqrt{-g}\left(R -2 \nabla^\mu\varphi\nabla_\mu\varphi +\lambda^2 f(\varphi) {\cal R}_{GB}\right),
\end{eqnarray}
where the scalar field $\varphi$ is coupled to the Gauss-Bonnet term ${\cal R}_{GB} \equiv R^2+R_{\mu\nu\rho\sigma}R^{\mu\nu\rho\sigma}-4R_{\mu\nu}R^{\mu\nu}$ through the Gauss-Bonnet coupling constant $\lambda$ and the coupling function $f(\varphi)$. As our main goal in this work is to search non-spontaneous scalarization of ECOs, we choose the coupling function to satisfy two following conditions
\begin{equation}
	\frac{df}{d\varphi}(\varphi=0) = 0, \quad \frac{d^2 f}{d\varphi^2}(\varphi=0) = 0. \label{TwoConditions}
\end{equation}
With the first condition, the theory admits GR vacuum solutions with vanishing scalar field---the Kerr metric. In this work, following the approach proposed in \cite{Maggio:2017ivp}, we consider a spherically symmetric horizonless compact object whose exterior geometry can be well described by the Schwarzschild metric
\begin{equation}
	ds^2 = - g(r) dt^2 + \frac{1}{g(r)} dr^2 + r^2 \left(d\theta^2 + \sin^2 \theta d\phi^2 \right),\quad r>r_s,  \label{Metric}
\end{equation}
where $g(r) = 1 - \frac{2 M}{r}$ with $M$ being the mass of the object. The object surface locates at $r=r_s$ which should be outside the would-be horizon $r =r_h = 2 M$. On this background, dynamics of the scalar field outside the object is governed by the modified Klein-Gordon equation
\begin{equation}
	\nabla_\mu \nabla^\mu \varphi= - \frac{\lambda^2}{4} \frac{d f(\varphi)}{d\varphi} {\cal R}_{GB}, \label{ScalarEq1}
\end{equation}
with ${\cal R}_{GB} = \frac{48 M^2}{r^6}$. From it, one can see that the scalar field acquires an effective mass square $m_{\rm eff}^2 = - \frac{\lambda^2}{4} \frac{d^2 f(\varphi)}{d\varphi^2} (\varphi=0) {\cal R}_{GB}$, which vanishes identically with the second condition of Eq. (\ref{TwoConditions}). This thus excludes the occurrence of the tachyonic instability, and so the scalarization, if exists, will be non-spontaneous.

Taking into account the conditions (\ref{TwoConditions}) and also the possible existence of stable scalar clouds, we consider the coupling function to take an exponential form
\begin{align}
	f(\varphi) = \frac{1}{4\kappa} \left(1 - e^{-\kappa \varphi^4}\right),\label{CouplingFunction}
\end{align}
with $\kappa$ being a parameter. This kind of coupling function has also been considered to discuss non-spontaneous scalarization of BHs in sGB \cite{Doneva:2021tvn}.

\section{Nonlinear instability}

In this section, we will first study the wave dynamics of the scalar field on the background (\ref{Metric}). As will show later, with this kind of coupling function (\ref{CouplingFunction}), the scalar field will experience a kind of nonlinear instability.

For simplicity, we assume the scalar field perturbation to be spherical symmetric, $\varphi = \varphi(t, r)$. In tortoise coordinate $dx \equiv \frac{dr}{g(r)}$, the scalar field equation (\ref{ScalarEq1}) becomes
\begin{equation}
	- \partial_t^2 \varphi + \partial_x^2 \varphi + \frac{2 g(r)}{r} \partial_x \varphi + \frac{\lambda^2 g(r)}{4} \frac{d f(\varphi)}{d\varphi} {\cal R}_{GB} =0, \label{ScalarPerturbationEq}
\end{equation}
which can be solved numerically by adopting the method of line \cite{schiesser2012numerical}. To solve the equations, physical boundary conditions are needed. At the object surface $r=r_s$, there are usually two boundary conditions considered, the Dirichlet and Robin boundary conditions. We will give more comments on the two boundary conditions in the last section. Let us first consider a Dirichlet boundary condition, so the physical boundary conditions we need to impose are that the scalar field vanishes at the object surface and is outgoing at infinity.

\begin{figure}[!htbp]
	\includegraphics[width=0.45\textwidth]{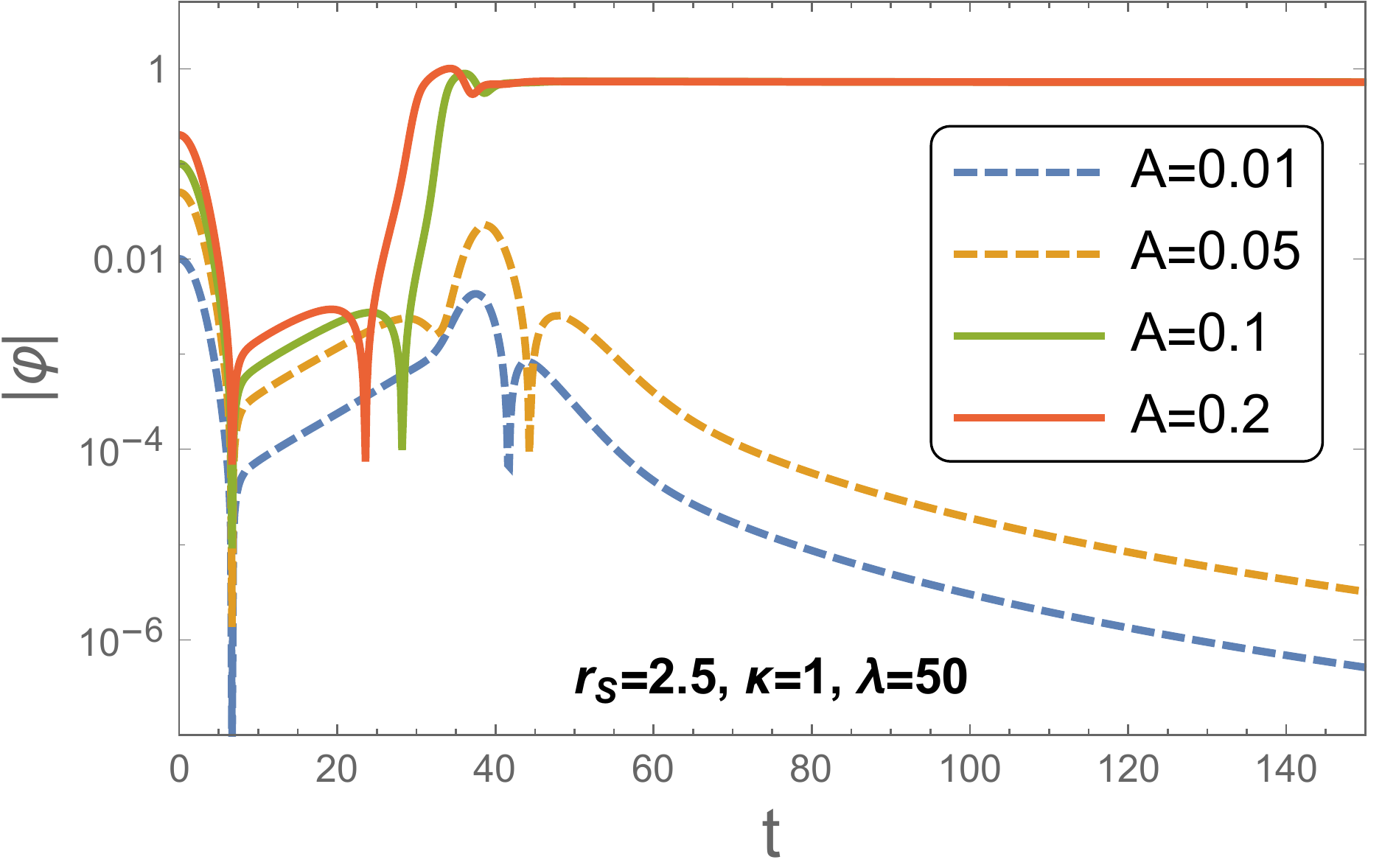}\quad
	\includegraphics[width=0.45\textwidth]{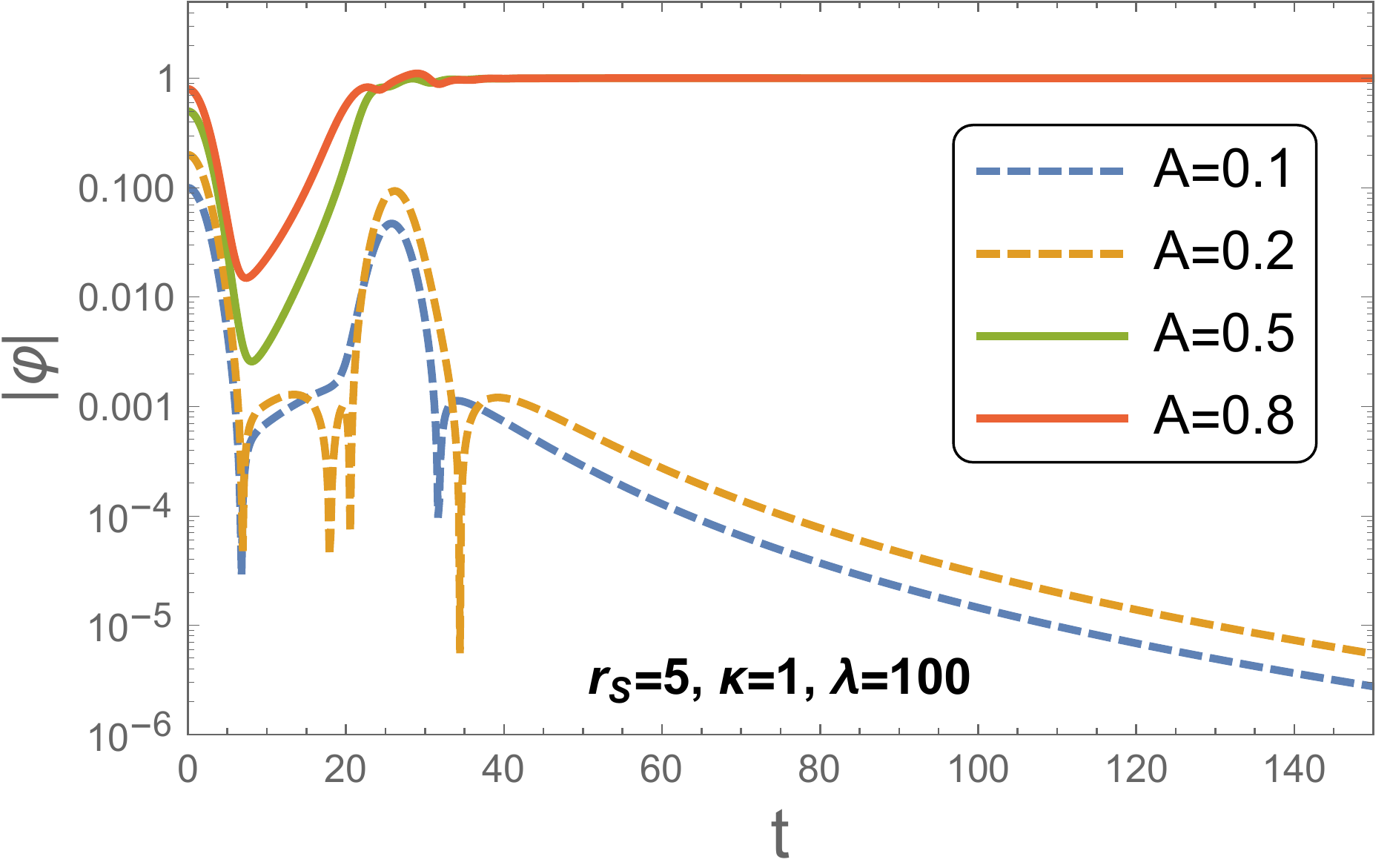}
	\caption{Time evolution of the scalar field perturbations on the spherically symmetric horizonless compact object with Dirichlet boundary condition. The perturbation is a Guassian pulse $\varphi(t=0, x) = A e^{-\frac{(x-x_c)^2}{2\sigma}}$ with $x_c=20$ and $\sigma=2$. Signal is extracted at $x=20$. We fix $M=1$ and all quantities are measured in units of $M$.} \label{TimeEvolutionDirichlet}
\end{figure}

We consider the initial perturbation to be a time-symmetric Gaussian pulse
\begin{equation}
	\varphi(t=0, x) = A e^{-\frac{(x-x_c)^2}{2\sigma}}, 
\end{equation}
with $x_c = 20M,\sigma = 2 M$ and $A$ the perturbation amplitude. There are four free parameters in the model, $\{M, r_s, \kappa, \lambda\}$. We fix $M=1$ and so all quantities are measured in units of it. By performing time evolution of the scalar field perturbation, we found that, depending on values of the parameters, nonlinear instability may be triggered. In Fig. \ref{TimeEvolutionDirichlet}, two typical examples are given. From the figure, one can see that the occurrence of the instability depends on the amplitude of the perturbation: When the amplitude $A$ is small, the perturbation will exhibit a decaying late-time tail; While $A$ becomes larger, the nonlinear term of $\varphi$ in the equation enters the game and the scalar field finally settles down to a equilibrium state indicating the formation of scalar cloud. This kind of instability, named as nonlinear instability, has already been observed in the BH scenario \cite{Doneva:2021tvn}.

\section{Scalar clouds with Dirichlet boundary condition}

In the last section, we have observed that the scalar field may experience a nonlinear instability, which is a signal of the formation of the scalar cloud. In this section, we will construct the scalar cloud explicitly. For simplicity, we consider the scalar cloud outside the object to be static and spherical, i.e., $\varphi=\varphi(r)$. After substituting the metric (\ref{Metric}) into (\ref{ScalarEq1}), the scalar field equation becomes
\begin{equation}
	\varphi''(r) + \left[\frac{2}{r} + \frac{g'(r)}{g(r)}\right] \varphi'(r) + \frac{\lambda^2}{4 g(r)} \frac{d f(\varphi)}{d\varphi} {\cal R}_{GB} = 0. \label{ScalarEq2}
\end{equation}
To solve the above equation, physical boundary conditions are needed. Assuming that a Dirichlet boundary condition at the object surface and the scalar field outside the object is regular and bounded, we have the boundary conditions
\begin{equation}
	\varphi (r_s) = 0,\quad \varphi (\infty) = 0. \label{BnyConditions}
\end{equation}

Now, the model contains four free parameters $\{M, r_s, \lambda, \kappa\}$. We can directly integrate the scalar field equation (\ref{ScalarEq2}) from the object surface to infinity for fixed values of these parameters. The number of free parameters can be reduced by noting that the scalar field equation (\ref{ScalarEq2}) possess two scaling symmetries
\begin{equation}
	r \rightarrow a r, \quad M \rightarrow a M,  \quad \lambda \rightarrow a \lambda, \label{Symmetry1}
\end{equation}
and
\begin{align}
	\varphi \rightarrow b \varphi, \quad \lambda \rightarrow b^{-1} \lambda, \quad \kappa \rightarrow b^{-4} \kappa, \label{Symmetry2}
\end{align}
where $a, b$ are arbitrary scaling parameters. With the two symmetries, it is convenient for numerical calculations to fix $M=1$ and $\varphi_1 \equiv \varphi'(r_s) = 1$, where $\varphi_1$ is the first radial derivative of the scalar field at the surface. So finally, there are left only three free parameters $\{r_s, \lambda, \kappa\}$. For a given $\kappa$ and $r_s$, the solution is determined uniquely by $\lambda$. However, not every value of $\lambda$ will produce a bounded solution that satisfies $\varphi(\infty) = 0$. Ony for certain  discrete values of $\lambda$ can produce bounded solutions. For a fixed $\kappa$ and $r_s$, we found an infinite countable set of the coupling constant, $\{\lambda(\kappa,r_s;n)\}_{n=0}^{n=\infty}$, which can support the bounded scalar clouds. Here a larger integer $n$ labels a larger $\lambda$.

\begin{figure}[t]
	\includegraphics[width=0.45\textwidth]{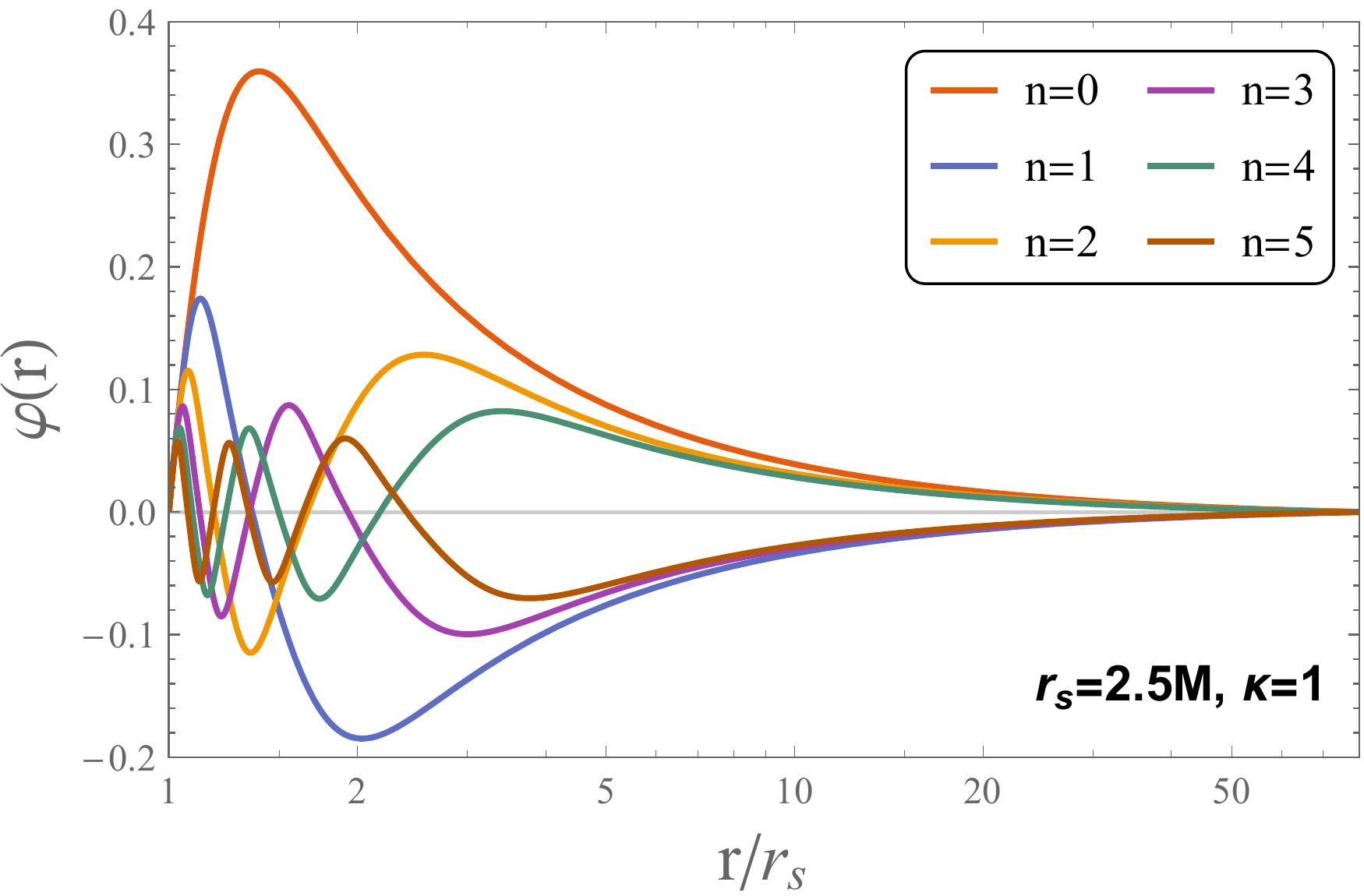}\quad
	\includegraphics[width=0.45\textwidth]{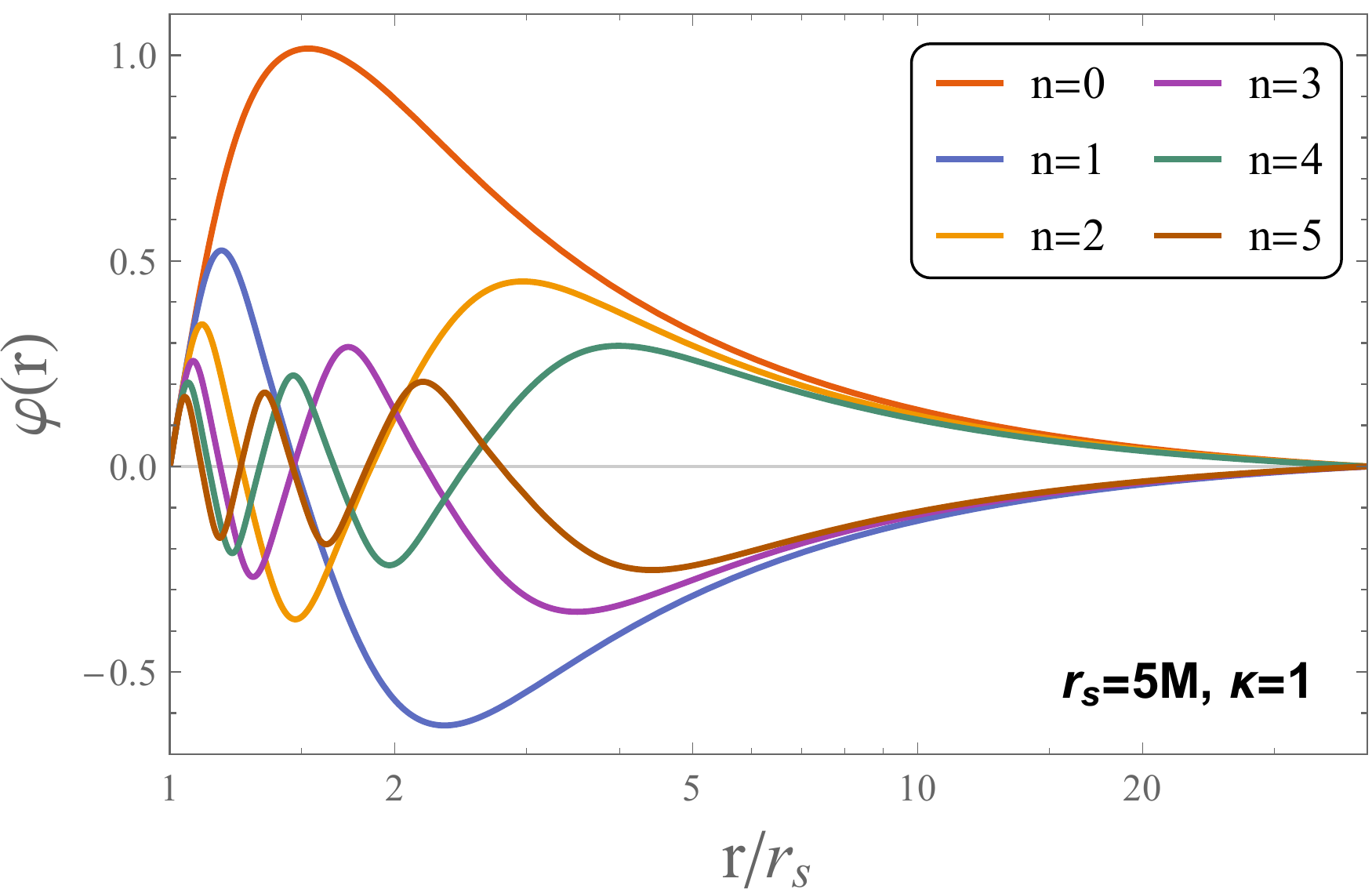}
	\caption{Configurations of the scalar clouds with Dirichlet boundary condition for various coupling constant $\{\lambda(\kappa,r_s;n)\}_{n=0}^{n=5}$. $n$ labels the number of nodes that the solution possesses between the object surface and radial infinity. We fix $\kappa=1$, $r_s=2.5M$ (left panel) and $r_s=5M$ (right panel).} \label{PhiFigDirichlet}
\end{figure}

\begin{figure}[t]
	\includegraphics[width=0.8\textwidth]{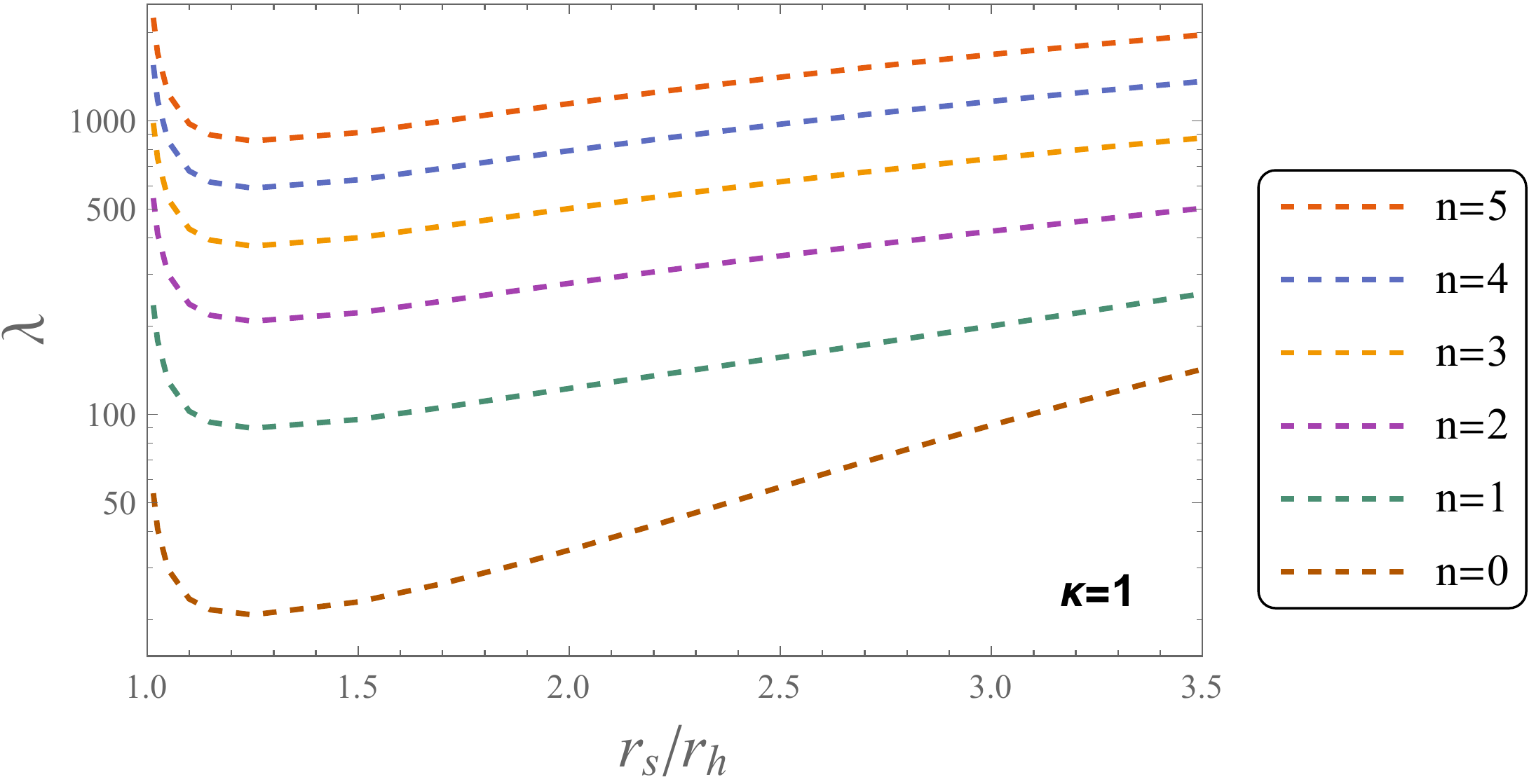}
	\caption{$\{\lambda(\kappa,r_s;n)\}_{n=0}^{n=5}$ as a function of $r_s$ for Dirichlet boundary condition. $r_h = 2 M$ is the would-be horizon.} \label{rslambdaFigDirichlet}
\end{figure}

In Fig. \ref{PhiFigDirichlet}, we show configurations of the scalar clouds for $\{\lambda(\kappa=1,r_s;n)\}_{n=0}^{n=5}$  with $r_s=2.5 M$ and $r_s=5M$ as examples. From the figure, one can see that the scalar clouds exhibit a standing-wave-like profile with $n$ indicating the number of nodes that the solution has between the object surface and radial infinity. Moreover, radial oscillations of the scalar clouds are concentrated in the region near the object surface. As the object radius $r_s$ increases, the oscillations move outward.

To survey the relation between $r_s$ and the coupling constant $\lambda$, we plot $\{\lambda(\kappa,r_s;n)\}_{n=0}^{n=5}$ as a function of $r_s$ in Fig. \ref{rslambdaFigDirichlet}. From the figure, one can see that, for each $n$, $\lambda(\kappa,r_s;n)$ is a convex function of $r_s$ with a minimum value $\lambda(\kappa,r_s;n)_{min}$. It is interesting to note that for any $n$, the minimum value appears at $r_s \approx 2.5 M$, which is close to but behind the photon sphere of the object. This suggests that scalar clouds are most likely to form when the object is very compact and has radius about $r_s \approx 2.5 M$. When $\lambda<\lambda(\kappa,r_s;n=0)_{min} \approx 21.45$, no scalar cloud can be supported for any $r_s$. When $\lambda(\kappa,r_s;n=0)_{min} \leq \lambda < \lambda(\kappa,r_s;n=1)_{min}$, only fundamental solution with $n=0$ exist. As $\lambda$ increases, excited solutions with higher $n$ emerge. Moreover, as $r_s$ approaches the would-be horizon $r_h=2 M$, $\{\lambda(\kappa,r_s;n)\}$ show a divergent behavior, which implies that ultra-compact objects are hard to support scalar clouds. 

\section{Scalar clouds with Robin boundary condition}

In the above, we have shown that compact objects can support scalar clouds with Dirichlet boundary condition. Another boundary condition usually considered is the Robin boundary condition \cite{Maggio:2017ivp}
\begin{equation}
	\frac{d(r \varphi)}{dr} (r_s) = 0.
\end{equation}

\begin{figure}[!htbp]
	\includegraphics[width=0.8\textwidth]{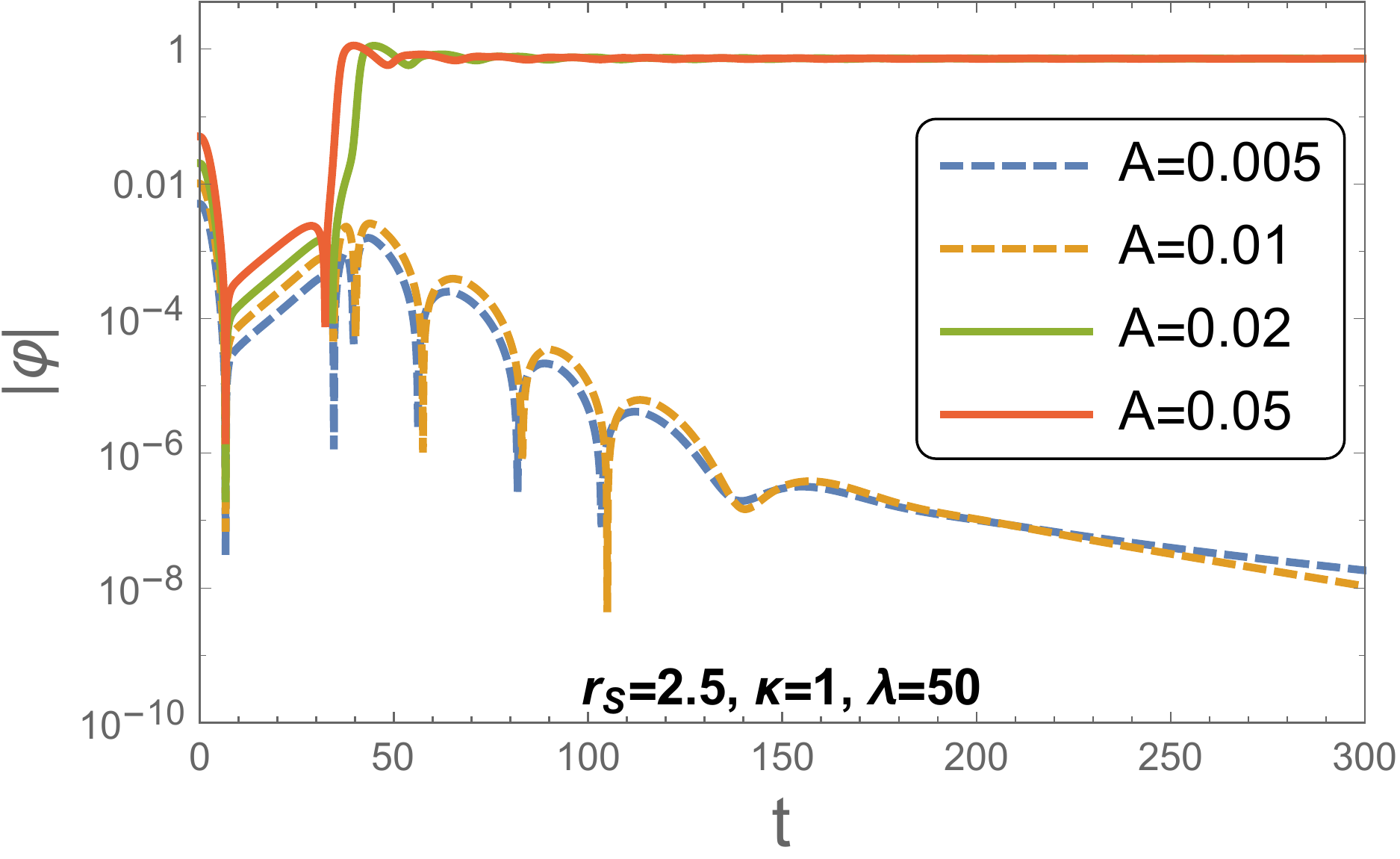}
	\caption{Time evolution of the scalar field perturbations on the spherically symmetric compact object with Robin boundary condition. The perturbation is a Guassian pulse $\varphi(t=0, x) = A e^{-\frac{(x-x_c)^2}{2\sigma}}$ with $x_c=20$ and $\sigma=2$. Signal is extracted at $x=20$. We fix $M=1$ and all quantities are measured in units of $M$.} \label{TimeEvolutionRobin}
\end{figure}

As in the case with Dirichlet boundary condition, the scalar field perturbations may also trigger nonlinear instability, as demonstrated in Fig. \ref{TimeEvolutionRobin}. Compared to the Dirichlet case in Fig. \ref{TimeEvolutionDirichlet}, we can see that, for small $A$, the perturbation experiences a longer ringdown phase before the late-time decaying.

\begin{figure}[!htbp]
	\includegraphics[width=0.45\textwidth]{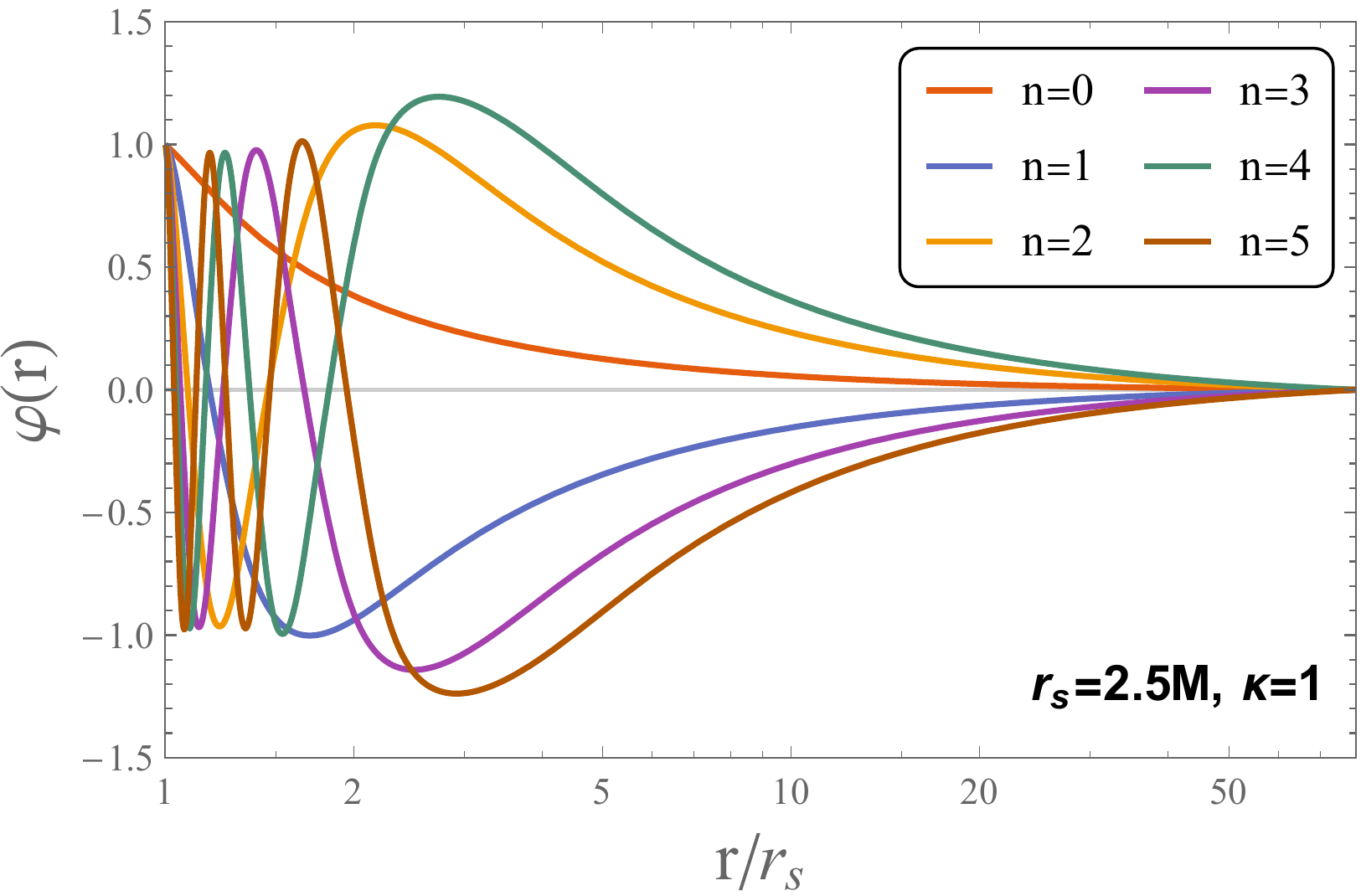}\quad
	\includegraphics[width=0.45\textwidth]{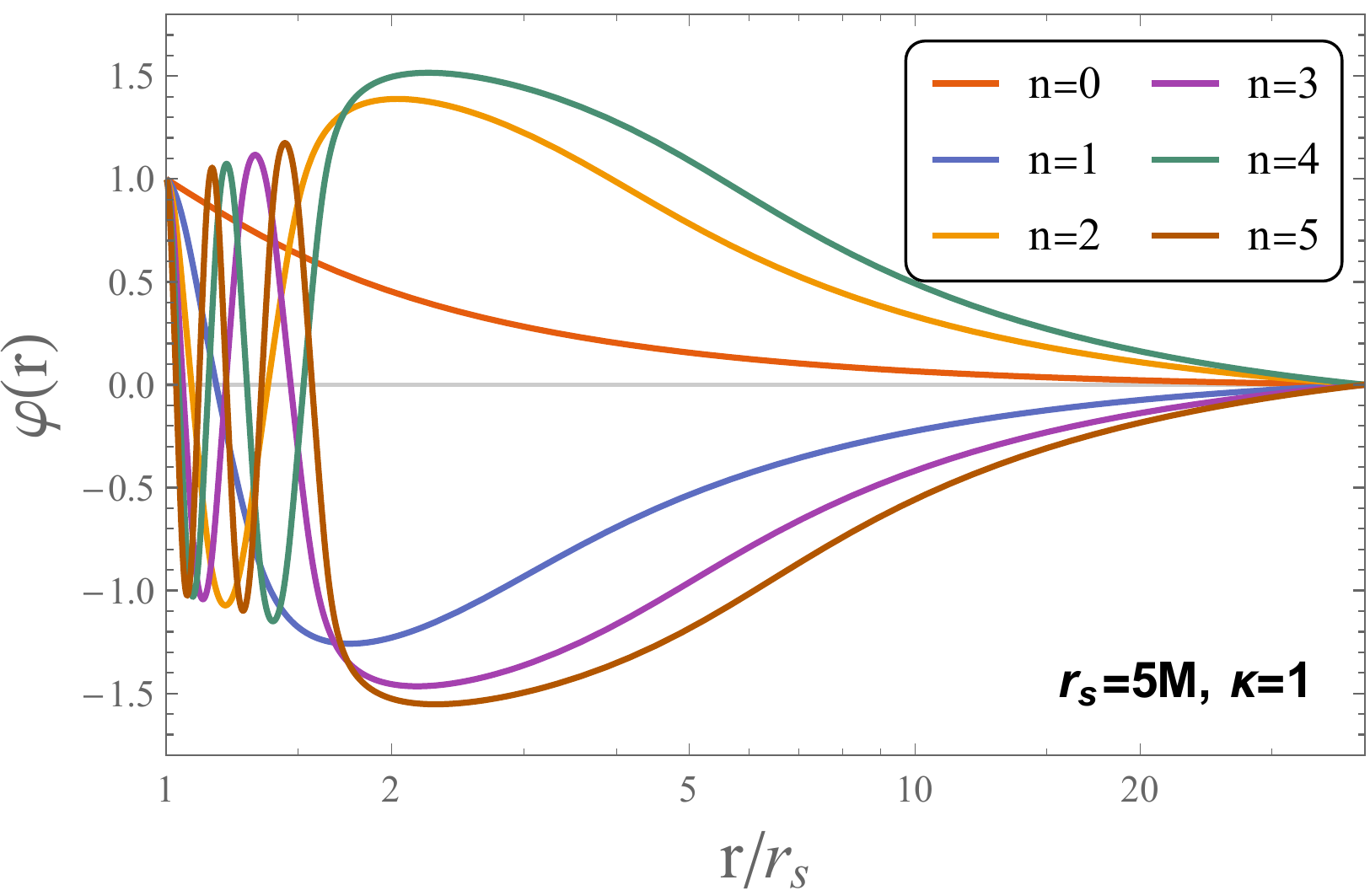}
	\caption{Configurations of the scalar cloud with Robin boundary condition for various coupling constant $\{\lambda(\kappa,r_s;n)\}_{n=0}^{n=5}$. $n$ labels the number of nodes that the solution possesses between the object surface and radial infinity. We fix $\kappa=1$, $r_s=2.5M$ (left panel) and $r_s=5M$ (right panel).} \label{PhiFigRobin}
\end{figure}

Similarly, we can construct the scalar cloud explicitly. In this case, it is convenient for numerical calculations to set $M=1$ and $\varphi(r_s) = 1$ with the two scaling symmetries (\ref{Symmetry1}) and (\ref{Symmetry2}). At infinity, we also have $\varphi(\infty) = 0$. As the case with Dirichlet boundary condition, for given values of parameters $\{\kappa, r_s\}$, there also exists an infinite countable set $\{\lambda(\kappa,r_s;n)\}_{n=0}^{n=\infty}$ which supports the scalar clouds. Samples are shown in Fig. \ref{PhiFigRobin}. Similar to the case with Dirichlet boundary condition, the scalar clouds also exhibit a standing-wave-like profile with the integer number $n$ labeling the number of nodes that they possess between the object surface and infinity. Compared to Fig. \ref{PhiFigDirichlet}, we can see that radial oscillations of the scalar clouds with Robin boundary condition are more concentrated in the region near the surface. And as the object radius $r_s$ increases, the oscillations move inward instead.

\begin{figure}[!htbp]
	\includegraphics[width=0.8\textwidth]{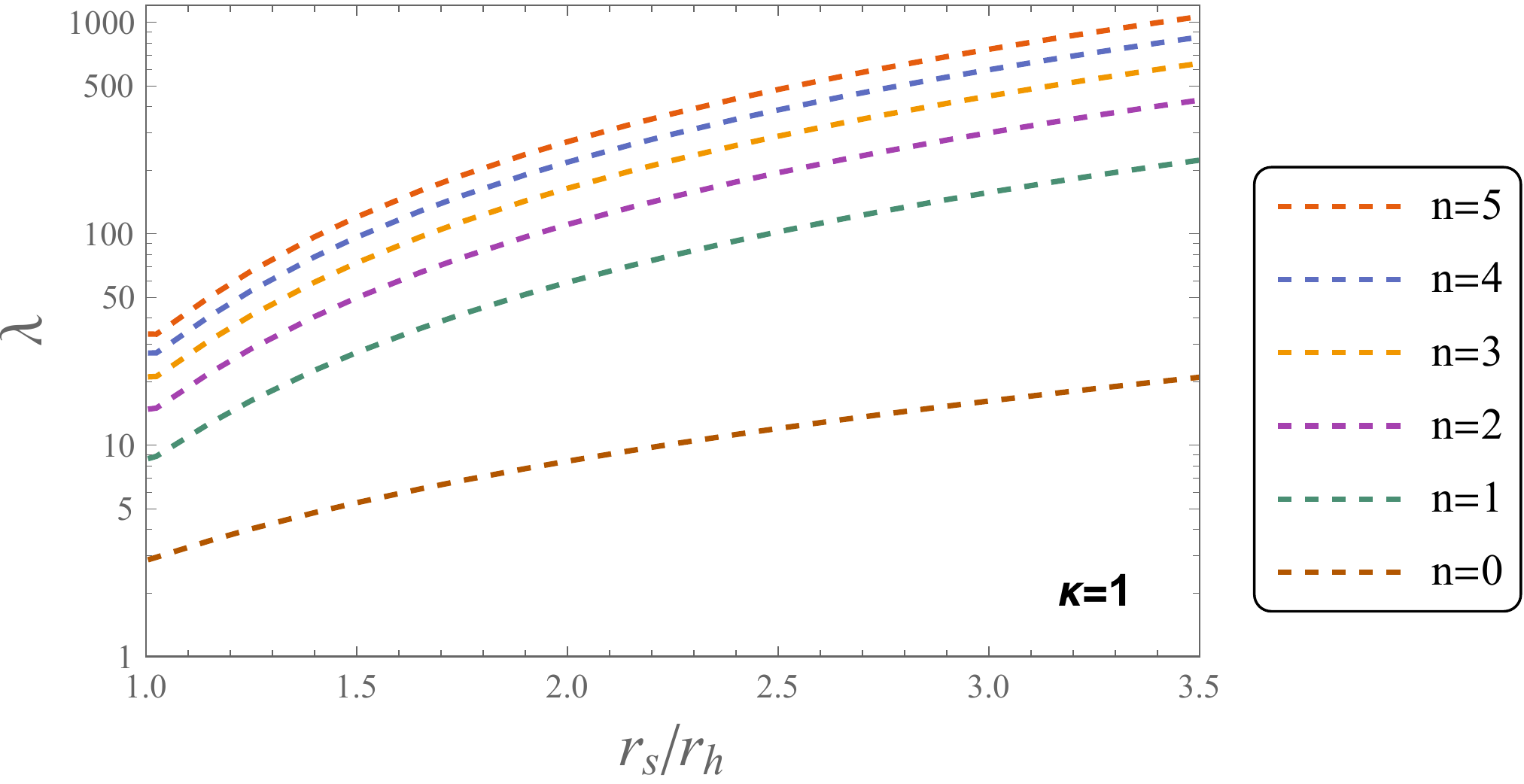}
	\caption{$\{\lambda(\kappa,r_s;n)\}_{n=0}^{n=5}$ as a function of $r_s$ for Robin boundary condition. $r_h = 2 M$ is the would-be horizon.} \label{rslambdaFigRobin}
\end{figure}

The relation between the object radius $r_s$ and the coupling constant $\lambda$ is shown in Fig. \ref{rslambdaFigRobin}. From the figure, one can observe one important feature different from that of the case with Dirichlet boundary condition: For each $n$, $\lambda$ is now a monotonically increasing function of $r_s$. This suggests that with the Robin boundary condition, scalarization becomes easier for more compact objects.

\section{Summary and Discussions}

In this work, we construct a novel type of scalar clouds for spherically symmetric horizonless compact objects in sGB theory. These scalar clouds are not formed due to tachyonic instability, but rather due to nonlinear instability, which requires a certain threshold of perturbation magnitude. We find that there exists an infinite countable set of coupling constants ${\lambda(\kappa,r_s;n)}_{n=0}^{n=\infty}$ (where $n$ is the number of nodes between the surface of horizonless compact objects and infinity) that can support the scalar clouds for a fixed radius $r_s$, given other parameter values. Moreover, we show that boundary condition (either Dirichlet or Robin) significantly affects the formation of scalar clouds. 

With Dirichlet boundary condition, $\lambda(\kappa,r_s;n)$ is a convex function of $r_s$, indicating that scalarization is most likely for compact objects with medium radii and becomes more difficult for ultra-compact and large ones.  However, with Robin boundary condition, $\lambda(\kappa,r_s;n)$ is a monotonically increasing function of $r_s$, suggesting that scalarization is easier for more compact objects. In both cases, there exists a threshold coupling constant $\lambda$ below which no scalar clouds can be supported. As $\lambda$ increases, fundamental solution with $n=0$ and excited solutions with higher $n$ emerge successively. 

In the context of Einstein's gravity with minimally coupled scalar field, these boundary conditions are commonly associated with a perfectly reflecting surface \cite{Maggio:2017ivp}. However, this interpretation is no longer valid in our theory due to the presence of a nonlinear term in $\varphi$, which prevents the scalar field perturbation equation (\ref{ScalarPerturbationEq}) from taking a wave-like form at the surface. Only when the perturbation is sufficiently small can we ignore the nonlinear term and regain this physical meaning. For general perturbations, the physical meaning of these boundary conditions is unclear and requires further investigation.

The coupling function we consider takes an exponential form with the exponent proportional to $\varphi^4$. We have also examined some other exponential forms, such as the one with exponent $\varphi^6$, and found similar instability and scalar clouds.

There are several possible extensions of this work. We work in the "decoupling limit" where the influence of the scalar clouds on the background geometry is neglected. This limit has shown its ability to capture the qualitative features of the fully nonlinear dynamics \cite{Doneva:2021dqn,Doneva:2021dcc}. However, it would be interesting to go beyond this limit and develop complete hairy solutions for further investigation. This would require specifying the explicit equation of state of the object and doing analysis model by model. In this paper we only consider spherically symmetric ECOs. A natural extension is to study the rotating case, where more than one physical mechanism may affect the formation of scalar clouds. Another intriguing question is whether other types of ECOs can support such scalar clouds.

\begin{acknowledgments}

	This work is supported by the National Natural Science Foundation of China (NNSFC) under Grant No 12075207.
\end{acknowledgments}

\bibliographystyle{utphys}%参考文献样式
\bibliography{Mylib}
\end{document}